# Node-Equivariant Message Passing for Efficient and Accurate Machine Learning Interatomic Potentials

Yaolong Zhang* and Hua Guo

Department of Chemistry and Chemical Biology, Center for Computational Chemistry, University of New Mexico, Albuquerque, New Mexico 87131, USA

* Corresponding author, email: ylzhangch@unm.edu

**Abstract**

Machine learned interatomic potentials, particularly equivariant message-passing (MP) models, have demonstrated high fidelity in representing first-principles data, revolutionizing computational studies in materials science, biophysics, and catalysis. However, these equivariant MP models still incur substantial computational and memory needs due to their expensive tensor product operations over edge space, significantly limiting their applicability in large-scale or long-time simulations. In this work, we propose a node-equivariant MP (NEMP) framework that performs equivariant operations between the central node and a virtual summed node encoding structure information of its neighbors. Crucially, NEMP maintains comparable or even superior accuracy across diverse test systems—including molecules, extended systems, and universal potential benchmarks—while achieving 1-2 orders of magnitude reduction in memory and computational costs compared to edge equivariant MP models. In fact, NEMP reaches computational efficiency comparable to that of local descriptor-based models, and enabling previously inaccessible large-scale simulations.

**Introduction**

Molecular dynamics (MD) is an essential tool for investigating both equilibrium and non-equilibrium properties of systems in gaseous and condensed phases. The reliability of such simulations naturally depends crucially on the accuracy of the interaction potential, but it is also important to realize that the system size and timescale that can be practically realized by such simulations are limited by the computational efficiency of these potentials. Recent advances in machine learning (ML) have revolutionized this field by representing first-principles data with efficient and high-fidelity machine-learned interatomic potentials (MLIPs)[1-33], enabling applications across various fields of chemistry, physics, and materials science[17, 32, 34].

A widely successful class of MLIPs builds upon the atomistic neural network (NN) framework introduced by Behler and Parrinello in 2007, which decomposes the total energy into atomic contributions dependent on local atomic environments within a cutoff radius[1]. These environments are encoded using many-body atomic descriptors that preserve translational, rotational and permutational symmetries. In early models, only two and three-body terms are included[1, 11], based on the "near-sightedness" assumption. However, such local descriptors are insufficiently sensitive to configurations with long-range differences. Studies have shown that increasing the body order of interactions in the descriptors can enhance model accuracy[13, 35-37]. To systematically incorporate higher-order terms, several approaches have been proposed, including the Moment Tensor Potential (MTP)[5] and the Atomic Cluster Expansion (ACE)[12]. However, numerical costs increase rapidly with the rising body order, which significantly limits the practical use of explicit higher-order terms in large systems.

An alternative and increasingly popular strategy for enhancing expressiveness is the message-passing NNs (MPNNs), which describe atomic environments by iteratively exchanging information between atoms (nodes) through connections between them (edges) in molecular graphs[7, 10, 20, 21, 23-27, 29, 38, 39]. This end-to-end learning framework is capable of implicitly capturing many-body interactions and some non-local effects beyond the cutoff radius[23]. Consequently, they are generally more efficient than NNs with explicit high body-order terms in improving the description of local structures. Early MPNN models[7, 10, 40] were based on two-body message functions, but subsequent studies showed that explicitly incorporating three-body (or higher-order) interactions into the message function allows for a more effective discrimination of local structures, which is crucial to increase the expressiveness of the MPNN model[21, 23, 41]. These strategies only pass symmetry-invariant scalar features and can be categorized as invariant MPNNs. Although invariant MPNNs can significantly improve model accuracy[21, 23, 42], they lack directional coupling between different atomic environments, which can lead to failure in certain systems where such interactions are critical[14, 27, 43]. This drawback can be easily remedied by passing equivariant features[24-26, 29, 39, 43-51]. These equivariant MPNNs extend the traditional paradigm by learning and exchanging equivariant features that transform predictably under symmetry operations. Indeed, models like NequIP[25], Allegro[29] and MACE[26] employ equivariant tensor products via the Clebsch-Gordan coefficients, leading to progressively and explicitly increased body-order interactions with each MP layer.

Until now, all equivariant MPNNs perform equivariant message passing along edge connections, where tensor products are computed between edges and their corresponding nodes. Hence, we refer to such an approach as edge-equivariant message passing (EEMP). While EEMPs have achieved high fidelity across diverse benchmarks[25, 26, 29, 52], they require an enormous computational overhead and memory requirement. Moreover, MPNNs face inherent limitations in parallelization efficiency due to massive inter-environment communication. This constraint limits their scalability to moderate-scale systems (for example, simulating liquid water comprising thousands to tens of thousands of atoms on an NVIDIA A100 GPU with 80 GB of memory[39, 53]). Although some equivariant MPNNs reduce costs by replacing tensor products with direct summations of equivariant features[25, 43, 51], such simplifications often compromise model accuracy.

In this work, we introduce a novel equivariant MPNN framework that leverages node-equivariant message passing (NEMP) to improve computational efficiency without sacrificing accuracy. Our key innovation lies in the reformulation of equivariant message passing in the node space, which are significantly smaller than the edge space. Rather than computing tensor products between every edge and corresponding neighbor node, we construct an expressive equivariant framework through a tensor product between the central node and a virtual summed node, where the neighbor features are summed with edge-dependent, highly expressive coefficients.

These coefficients and features are iteratively refined during MP, preserving physical symmetries while maintaining or even increasing the model's expressive power. As a result, the expensive scaling in the equivariant MP step is reduced from with the number of neighboring pairs to with the number of central atoms. Extensive evaluations across molecular systems, gas-solid interfaces, liquids and solids demonstrate that our framework achieves high fidelity in representing the first-principles data while improving computational and memory efficiency by 1-2 orders of magnitude. This improvement allows not only reduction of simulation time, but also enable simulations of much larger systems.

**Results**

**NEMP framework**

Equivariant MPNNs have gained significant attention in atomic-scale ML for their remarkable fidelity and data efficiency[34]. In these models, molecular structures are considered as graphs, where atoms are treated as nodes and edges are defined as the connections between the central atom and its neighbors within a specified cutoff radius. The MP is realized iteratively in layers, in which each node gathers information from its surrounding atoms, typically encoded with symmetric features, and combines them through a learned aggregation function, which is termed massage. The aggregated information is then used to pass the message by updating the node features. Through repeated layers, the network captures increasingly complex correlations and interactions across the molecular graph. The value of equivariant MPNNs lies in their ability to couple these node states through the tensor product of equivariant features of the central atom and its neighbors. This mechanism not only preserves equivariance but also introduces explicit incorporation of higher-order interactions, in contrast to invariant MPNNs, which capture such interactions only implicitly[23, 54].

The most critical component of equivariant MPNNs lies in how they perform equivariant message passing. In EEMP models, equivariant messages are passed by coupling equivariant features of each edge with those of the corresponding neighbor node, as illustrated in Fig. 1(a). This strategy results in two tensors of shape ($N_{atom} \times N_{neigh} \times N_{lmc} \times N_c$), where $N_{atom}$ is the number of atoms in the system, $N_{neigh}$ is the number of neighbors, $N_{lmc}$ denotes the number of angular momentum combinations, and $N_c$ is the number of channels (*vide infra*). In particular, $N_{lmc}$ increases rapidly with the maximum angular momentum, exhibiting cubic scaling. This results in the construction of extremely large tensors and, consequently, significant computational cost in both time and memory. As a result, the tensor product becomes the rate-determining step, severely limiting the applicability of such models to large systems or long-time simulations. Below, we propose the NEMP to avoid some of the shortcomings of EEMP.

**NEMP Architecture**

The NEMP architecture is shown in Fig. 1(b). We first need to generate a representation for each unique pair of atoms to enable its application in describing multi-element systems, which is crucial for building a universal potential across various elements. This can be achieved through an embedding NN, denoted as $F_{emb}$, which takes the one-hot encoding combination of a pair of elements $\{Z_i, Z_j\}$ as input[29]:

$$\boldsymbol{\sigma}_{ij} = F_{emb}\left(\{Z_i, Z_j\}\right). \tag{1}$$

Here, $Z_i$ and $Z_j$ represent the element types of the central and neighbor atoms, respectively. The output of this embedding depends only on the combination of element types and is used to generate parameters that similarly depend only on $\boldsymbol{\sigma}_{ij}$ in subsequent processing:

$$\mathbf{c}(\boldsymbol{\sigma}_{ij}) = F_{coeff}\left(\boldsymbol{\sigma}_{ij}\right). \tag{2}$$

For the NEMP description of the local edge environment within a cutoff radius of the $i$th central atom, Bessel functions with optimizable parameters are used for the radial channels:

$$R_n(r_{ij}) = \frac{\sin\left(c_n^R\left(\boldsymbol{\sigma}_{ij}\right) r_{ij}\right)}{c_n^R\left(\boldsymbol{\sigma}_{ij}\right) r_{ij}} f_{cut}(r_{ij}; r_c). \tag{3}$$

Here, $j$ denotes the surrounding atoms within the cutoff radius $r_c$, enforced by a polynomial cutoff function[55] $f_{cut}(r_{ij}; r_c)$. $c_n^R\left(\boldsymbol{\sigma}_{ij}\right)$ is the output of the NN ($F_{coeff}$) defined in the Eq. (2) for the radial functions (denoted by superscript $R$), with $n$ ranging from 1 to $N_c$, where $N_c$ is the number of channels (radial functions) in the initial layer.

Before detailing the specific operations in each MP layer, we first introduce the notational conventions used throughout the text, where bold symbols denote vectors with channel dimensions. The notations "$I$" and "$J$" refer to the central node ($i$) and a neighboring node ($j$), respectively, while "$IJ$" denotes the edge connecting them. The first Greek subscript indicates the specific type of feature, which include edge coefficients (ECs) ($\chi$), equivariant ($\psi$), and invariant ($\rho$) features. Finally, the numeral superscript denotes the layer, while the absence of a superscript indicates that the quantity is an intermediate result within the layer that is not carried over to the next MP layer. Finally, the Greek superscript relates to the feature character.

By applying an NN to the combination of radial functions in Eq. (3) and the element-dependent coefficients $\mathbf{c}^{\chi}(\boldsymbol{\sigma}_{ij})$ (Eq. (2)), we generate the environment dependent ECs of the initial layer:

$$\mathbf{IJ}_{\chi}^{0} = F_{edge}\left(\left\{\mathbf{c}^{\chi}(\boldsymbol{\sigma}_{ij}), \mathbf{R}(r_{ij})\right\}\right). \tag{4}$$

The corresponding node-equivariant features ($\mathbf{I}_{\psi,lm}^{0}$) for the central atom $i$ can be obtained by directly summing the contributions of their neighboring atoms denoted by $j$:

$$\mathbf{I}_{\psi,lm}^{0} = \sum_{j} \mathbf{IJ}_{\chi}^{0,\psi} Y_{lm}(\mathbf{r}_{ij}). \tag{5}$$

Here, $Y_{lm}(\mathbf{r}_{ij})$ are the spherical harmonics, $\mathbf{IJ}_{\chi}^{0,\psi}$ are the initial layer ECs defined in Eq. (4) related to the node-equivariant features. On the other hand, the node-invariant features ($\mathbf{I}_{\rho}^{0}$) in the initial layer can be generated by summation over the ECs:

$$\mathbf{I}_{\rho}^{0} = \sum_{j} \mathbf{IJ}_{\chi}^{0,\rho}. \tag{6}$$

For the first MP layer, edge-invariant features are derived from node-equivariant features of the initial layer defined in Eq. (5):

$$\mathbf{IJ}_{\rho}^{1} = \sum_{l,m}\left(\mathbf{I}_{\psi,lm}^{0} \mathbf{IJ}_{\chi}^{0,\rho_i} Y_{lm}(\mathbf{r}_{ij}) + \mathbf{J}_{\psi,lm}^{0} \mathbf{IJ}_{\chi}^{0,\rho_j} Y_{lm}(\mathbf{r}_{ij})\right), \tag{7}$$

These newly computed edge-invariant features $\mathbf{IJ}_{\rho}^{1}$ are then combined with the edge-invariant features from the initial layer $\left\{\mathbf{c}^{\rho}(\boldsymbol{\sigma}_{ij}), \mathbf{IJ}_{\chi}^{0}\right\}$ to produce refined edge-invariant features for the current layer $\mathbf{IJ}_{\rho}^{1} = \left\{\mathbf{c}^{\rho}(\boldsymbol{\sigma}_{ij}), \mathbf{IJ}_{\chi}^{0,\rho}, \mathbf{IJ}_{\rho}^{1}\right\}$. Using these updated features, the ECs ($\mathbf{IJ}_\chi^1$) for the first layer are generated using the same form as that in the initial layer as shown in Eq. (4). With these ECs, we perform a tensor contraction between the weighted spherical harmonic expansion and the node-equivariant features from the previous layer to obtain the new node-invariant features ($\mathbf{I}_\rho^1$):

$$\mathbf{I}_\rho^1 = \sum_{l,m} \mathbf{I}_{\psi,lm}^0 \sum_j \mathbf{IJ}_\chi^{1,\rho} Y_{lm}(\mathbf{r}_{ij}). \tag{8}$$

To overcome the bottleneck associated with EEMP, as previously discussed, we design our MP scheme such that the tensor product is performed only over the node space. For each central node, we first obtain its equivariant features ($\mathbf{I}_{\psi,lm}$):

$$\mathbf{I}_{\psi,lm} = \sum_j \mathbf{IJ}_\chi^{1,\psi} Y_{lm}(\mathbf{r}_{ij}). \tag{9}$$

To acquire information outside the cutoff sphere, we introduce a virtual summed node (~), as illustrated in Fig. 1(a). The corresponding equivariant feature ($\tilde{\mathbf{I}}_{\psi,lm,n}$), can be obtained by aggregating or summing over its neighboring nodes

$$\tilde{\mathbf{I}}_{\psi,lm,n} = \sum_j IJ_{\chi,n}^{1,\tilde{\psi}} \sum_{\tilde{n}=1} \omega_{j,\tilde{n}n}^1 \mathbf{J}_{\psi,lm,\tilde{n}}^0. \tag{10}$$

The message passing is then realized by performing a tensor product between the node-equivariant features of the central and virtual nodes:

$$\mathbf{I}_{\psi,l_f m_f}^1 = \sum_{l_1,m_1,l_2,m_2} C_{l_1,m_1,l_2,m_2,l_f,m_f} \omega_{i,l_1,l_2,l_f}^1 \tilde{\mathbf{I}}_{\psi,l_1 m_1} \mathbf{I}_{\psi,l_2 m_2}, \tag{11}$$

where, $C_{l_1,m_1,l_2,m_2,l_f,m_f}$ are the corresponding Clebsch–Gordan coefficients. This strategy produces two tensors of shape ($N_{atom} \times N_{lmc} \times N_c$), without the $N_{neigh}$ dimension as in the EEMP model. As a result, it is significantly more efficient in terms of both time and memory, removing the bottleneck in EEMP.

The final node-equivariant features for the first layer are obtained through a ResNet-style connection[25, 56]:

$$\mathbf{I}^1_{\psi,lm,\tilde{n}} = \sum_{n=1} \tilde{\omega}^1_{i,\tilde{n}n} \mathbf{I}^0_{\psi,lm,n} + \sum_{n=1} \omega^1_{i,\tilde{n}n} \mathbf{I}^1_{\psi,lm,n}. \tag{12}$$

In Eqs. (10) and (12), $\tilde{\omega}$ represents different optimizable parameters, where the subscripts indicate their dependencies. For example, $\omega^1_{i,\tilde{n}n}$ in Eq. (10) denotes that the optimizable parameter depends on the element species ($j$) and the channel coupling ($\tilde{n}n$). Importantly, in Eqs. (10) and (12), we use element-dependent coefficients for basis contraction before further product or sum coupling. This approach facilitates information exchange between different channels, similar to basis contraction in quantum chemistry, and has been shown to enhance expressiveness[57].

Subsequently, the node-equivariant features are employed to generate the edge-invariant features using the same form in first layer as shown in Eq. (7), allowing the message-passing process to continue as shown in Fig. 1(b) and (c). After $T$ MP iterations, the refined edge-invariant features converge to a sufficiently expressive representation of the local edge environment, naturally producing fully resolved coefficients. Therefore, Eqs. (9) and (10) can be regarded as a fully inverting process (similar to solving a system of equations), meaning that if we have enough channels, we can recover all the original edge information summed in Eqs. (9) and (10). Therefore, our model, through the tensor product in Eq. (11), does not sacrifice expressive power, but with much lower computational cost and memory consumption than the EEMP.

Similarly, the node-invariant features develop a faithful representation of the entire system structure. The model aggregates (e.g., concatenates or integrates) node-invariant features of each layer and processes them through an NN to predict atomic energies. The total energy of the system is finally obtained by summing these atomic contributions $E=\sum_{i}^{N} F_{readout}\left(\{\mathbf{I}_{\rho}^{0}, \cdots, \mathbf{I}_{\rho}^{T}\}\right)$.

Furthermore, the sum operation can introduce additional interactions, as shown in Fig. 1(a) and (c). In the EEMP model, each edge-equivariant feature (spherical harmonics) only interacts with the corresponding neighbor's node-equivariant features. For example, using the local environment centered at atom $I$ depicted in Fig. 1(a), the interaction of an EEMP model is expressed as

$$\mathbf{IJ}_{\chi}(r_{ij})\mathbf{Y}(\mathbf{r}_{ij})\otimes \mathbf{J}_{\psi} + \mathbf{IK}_{\chi}(r_{ik})\mathbf{Y}(\mathbf{r}_{ik})\otimes \mathbf{K}_{\psi} + \mathbf{IH}_{\chi}(r_{ih})\mathbf{Y}(\mathbf{r}_{ih})\otimes \mathbf{H}_{\psi}, \qquad (13)$$

Here, $J$, $H$, and $K$ denote the neighboring nodes. Similarly, in Fig. 1(c), the interaction in the NEMP model is defined as follows:

$$\left(\mathbf{IJ}_{\chi}^{\psi}\mathbf{Y}(\mathbf{r}_{ij})+\mathbf{IK}_{\chi}^{\psi}\mathbf{Y}(\mathbf{r}_{ik})+\mathbf{IH}_{\chi}^{\psi}\mathbf{Y}(\mathbf{r}_{ih})\right)\otimes \left(\mathbf{IJ}_{\chi}^{\psi}\mathbf{J}_{\psi}+\mathbf{IK}_{\chi}^{\psi}\mathbf{K}_{\psi}+\mathbf{IH}_{\chi}^{\psi}\mathbf{H}_{\psi}\right), \qquad (14)$$

There are more direct interactions and many-body based coefficients in our model since coefficients in EEMPs only depend on distances between the central atom and its neighbors, while our NEMP model depends on iteratively refined many-body edge features. The ECs are learned from data to automatically determine interaction significance. Numerical results presented below show that our model can achieve state-of-the-art fidelity in many systems, while offering substantially higher efficiency and requiring fewer parameters (approximately 50k to 500k) compared with EEMP

models[25, 52, 58].

**Single-component systems**

**3BPA**

For MLIP models, extrapolation capability, which is the ability to accurately predict out-of-domain data, is a crucial measurement for assessing model performance. To evaluate the extrapolation capability of NEMP, we employ the flexible drug-like molecule 3-(benzyloxy)pyridin-2-amine (3BPA) dataset, sampled from MD simulations at 300, 600, and 1200 K[59]. The reference energies and forces were computed using density functional theory (DFT) with the $w$B97X functional and the 6-31G basis set, as described in ref. [59]. Following the same protocol, our model was trained with 500 structures from the 300 K dataset, which was randomly split into training and validation sets in a 9:1 ratio. In addition, configurations were sampled at 300, 600, and 900 K to construct three separate test datasets. Given the substantial energy range difference between the test datasets in high temperature (~4.5 eV) and the training dataset (1.1 eV), we adopt a linear mapping of edge-invariant features to ECs (Eq. (4)) in each MP layer to enhance transferability. The test error was evaluated on three distinct sets to assess both in-domain and out-of-domain accuracy.

Comparisons with several existing MLIP models, including ACE[59], SGDML[60], ANI-2x[61], CACE[39], NequIP[25, 26], Allegro[29], and MACE[26], are summarized in Table I. Notably, NEMP achieves superior fidelity compared to all models except MACE, for which the errors are practically comparable. This supports our claim that NEMP is effectively equivalent to the EEMP framework. We note in passing that MACE incorporates additional many-body interactions by leveraging node-equivariant features as a basis after each MP layer, enabling higher-order representations, so it is not surprising that it achieves better accuracy than other MLIPs. As demonstrated in examples discussed below, however, introducing an NN to map the edge-invariant features to ECs in each layer can mitigate this issue, yielding comparable or even better performance than MACE.

**Liquid water**

To validate the performance of the NEMP model on complex liquids, we benchmark it against ab initio reference data generated by Cheng et al.[35] The dataset comprises 1,593 structures of 64 water molecules in a periodic box, computed using DFT with the revPBE0-D3 functional. This level of theory has been demonstrated to provide a reliable description of water's structure and dynamics across a range of pressures and temperatures. In Table II, we compare the performance of several models, including (1) local descriptor-based atomic NNs, (2) invariant MPNNs, and (3) equivariant MPNNs. Among these, two and three-body descriptor-based NNs like the Behler-Parrinello NN (BPNN)[35] and the embedded atom neural network (EANN)[11, 62] give the largest errors, as expected. REANN (invariant MPNN)[57] significantly reduces the errors of local descriptor-based models, reaching comparable accuracy with equivariant models NequIP[25] and CACE[39]. Once again, MACE[58] demonstrated superior fidelity, achieving a root mean square error (RMSE) of 1.9 meV/atom for energy and 36.2 meV/Å for forces, outperforming other EEMP models. Notably, Our NEMP model, which incorporates an NN as the edge-feature update function $F_{edge}$, obtains the error to 2.3 meV/H$_2$O and 37.3 meV/Å while using significantly fewer parameters, as detailed in the Training Details section. This result supports our claim that introducing NN-mapped many-body ECs enhances the model's expressive power.

To further evaluate the reliability of the NEMP potential, we performed Nose-Hoover thermostat-based NVT MD simulations of the liquid water system consisted of 64 H$_2$O molecules in a periodic cubic box, equilibrated at 300 K. The simulations were carried out for 50 ps with a time step of 0.1 fs. Fig. 2 compares the oxygen-oxygen radial distribution function (RDF) obtained from the NEMP potentials with experimental results[63]. The close agreement between the NEMP predictions and experimental measurements further demonstrates the accuracy in reproducing the reference DFT energies and interatomic forces.

**Universal Potential Applications**

**ANI-1x**

After demonstrating the high fidelity of the NEMP model for single-component systems, we further evaluate its generalization ability for constructing universal potentials. To this end, we benchmark the model using the ANI-1x dataset—a dataset for training transferable universal potentials for organic molecules with H, C, N, and O elements[64]. Following the protocols adopted in MACE, we trained our model on a subset of ~490k data points (approximately 10% of the full dataset) and evaluated its performance on the COMP6 benchmark dataset[65]. As presented in Table III, the NEMP model outperforms all existing approaches (including ANI-1x[66], EEMP model MACE[58] and TensorNet[67]), achieving a total mean absolute error (MAE) 0.48 kcal/mol for energy and 0.39 kcal/mol/Å for forces. These results are consistent with the findings above, confirming that NN-mapped many-body ECs can enhance the model's expressive power. Notably, while the training data only included systems with ⩽50 atoms, the model successfully extrapolates to larger molecular systems (up to 300 atoms). These results highlight NEMP as a promising and scalable framework for developing general-purpose potentials.

**HME21**

To further test the performance of the NEMP model for materials science applications, we choose to work with the HME21 dataset[68], which includes regular and disordered crystals spanning 37 elements. This diverse chemical space makes HME21 an ideal benchmark for evaluating the generalizability of MLIPs for predicting structures and energies of materials. Accurately modeling systems with such a large elemental variety has long been considered a key challenge in MLIP construction[41, 50, 69-71]. For fair comparison, we use the same data split as in the original test. As shown in Table IV, NEMP achieves accuracy comparable to MACE[58] while outperforming NequIP[25, 58] and TeaNet[47, 58]. The consistently low errors confirm NEMP's ability to handle extensive elemental diversity without sacrificing precision in energy and force predictions.

**EMLP dataset**

The Element-based Machine Learning Potential (EMLP)[72] is a recent NequIP trained MLIP encompassing Ag, Pd, C, H, and O, designed for applications in heterogeneous catalysis and beyond. Comprising of 116,516 DFT-calculated data points generated through random chemical space exploration, the dataset prioritizes diverse local atomic interactions over extensive collections of structurally similar configurations to ensure generalizability. EMLP demonstrates accurate property predictions across solid, liquid, gas phases and gas-surface systems without requiring system-specific sampling. Perhaps most importantly, it demonstrates a superior ability in predicting chemical transformation barriers (including transition state geometries and energies), which is often lacking in other MLIPs. Furthermore, it has also been demonstrated to correctly describe liquid molecular systems with only limited data. The inherent complexity and demonstrated transferability of the EMLP dataset presents it as an excellent benchmark for evaluating performance of MLIPs.

Following the same training strategy, we developed an NEMP model using the EMLP dataset and evaluated its performance by comparing predicted energetic profiles for CO oxidation on various Pd surfaces against reference DFT calculations. As shown in Fig. 3, both models show good agreement with DFT results, with NEMP achieving comparable or even superior accuracy to EMLP in certain cases, consistent with our observations in other systems discussed above.

Beyond gas–surface systems, we further validated this universal potential through MD simulations of liquid ethanol. We carried out Nose–Hoover thermostated NVT simulations, with a system comprising 32 $CH_3OH$ molecules in a periodic cubic cell (side length: 12.93 Å) equilibrated at 300 K. The simulation was run for 50 ps with a time step of 0.1 fs. Fig. 4 compares the C–C, C–O, and O–O RDFs predicted by the NEMP potential with those from ab initio molecular dynamics (AIMD), EMLP, and experimental results[73]. The NEMP predictions are obviously in closer agreement with AIMD than those of EMLP, which underscores the accuracy of NEMP in reproducing reference energies and interatomic forces, further confirming its robust transferability across different phases and systems.

**Computational Efficiency and Scalability of NEMP**

By design, the NEMP represents significant advantages in computational efficiency over existing equivariant models. As displayed in Table I for the 3BPA molecule, NEMP reduces the computational cost by approximately 1-2 order of magnitude compared to EEMP baselines (MACE/NequIP/Allegro) while maintaining comparable accuracy.

To evaluate CPU scalability, we measured the computational cost per MD step in NVT simulations of liquid water (300 K) using the JAX_MD package[74]. Fig. 5(a) demonstrates linear scaling with system size on a single Intel® Xeon Gold 6438Y core. The NEMP potential shows marginally higher computational cost than BPNN[1, 62] (an efficient local-descriptor atomistic model). For a more detailed comparison, we reduced the number of hyperparameters in our model, creating an NEMP-small variant with RMSE of 3.0 meV/$H_2O$ for energy and 48.6 meV/Å for forces, maintaining accuracy comparable to EEMP models. This model achieves a speed of $2.2 \times 10^{-4}$ s/atom/step, representing approximately 2.5× acceleration over the standard NEMP and 2× faster performance than the efficient local-descriptor-based BPNN model. Notably, while the REANN (invariant MPNN) design theoretically promises higher speed than EEMP, both NEMP variants demonstrate similar performance in our CPU benchmarks.

For GPU benchmarks, memory emerges as a critical constraint due to hardware limitations. Our solution is a novel asynchronous architecture (Fig. 6) that decouples neighbor-list (NL) construction on the CPU from force evaluation on the GPU. The implementation combines JAX's non-blocking execution with a skin-algorithm cutoff buffer to create pipelined updates: the GPU computes forces using historical NLs (time $t$-$n$) while the CPU simultaneously generates new NLs based on current positions (time $t$) through optimized Fortran code, with asynchronous NL transfers to the GPU every $n$ steps. This design eliminates GPU waiting time by masking NL latency and overcomes the high memory consumption of NL calculations for general lattice parameters in JAX_MD required for just-in-time (JIT) compilation compatibility (limited to ~50k atoms), enabling nearly one million atoms simulations. This strategy can be readily adapted to other ML models for enhanced memory and computational efficiency for efficient GPU models.

Benchmarks on NVIDIA A800/H100 GPUs (Fig. 5(b)) demonstrate the computational and memory efficiency of our model. Comparison between Figs. 5(a) and 5(b) reveals GPU/CPU acceleration ratios that scale with system size, reaching a maximum of ~200× for systems exceeding ~10k atoms. NEMP-small successfully simulates nearly one million atoms within 80 GB memory, in contrast to EEMP models (MACE/NequIP/CACE) which are limited to 10k ~ 24k atoms with a speed around $0.9 \times 10^{-4}$ ~ $1.5 \times 10^{-4}$ s/atom/step on A100 GPUs with 80 GB memory. Notably, NEMP-small achieves 5× faster evaluation speeds than REANN. Compared to EEMP models, NEMP-small reduces computational costs by two orders of magnitude while improving memory efficiency by 30-70× on A800/A100 GPUs, further achieving 0.5 μs/atom/step on H100 GPUs. Benchmark results validate our model's high computational efficiency and low memory consumption.

It is important to note that this asynchronous strategy is most effective when the evaluation time of the ML model is comparable to, or shorter than, the overhead associated with CPU-GPU data transfer, or when using frameworks that require JIT compilation, such as JAX-MD, where constructing NLs for general lattice parameters incurs substantial memory overhead. In contrast, when the cost of force evaluation significantly exceeds these overheads, a conventional sequential setup, where NL construction and MD steps are performed entirely on either the CPU or the GPU, can also achieve near-optimal performance.

**Conclusions**

We present here a novel node-based equivariant MPNN framework that achieves high efficiency and accuracy in modeling MLIPs across diverse chemical systems. Central to our approach is the tensor product operation between node states and virtual summed node states, weighted by many-body ECs generated through NNs within each MP layer. This design significantly reduces computational costs while enhancing expressive power.

Benchmark results demonstrate that our model achieves accuracy comparable to or exceeding prevailing EEMP models across datasets designed for gas-phase molecules, extended systems, and universal potentials. MD simulations confirm 1~2 orders of magnitude acceleration relative to existing EEMP models while maintaining accuracy, achieving speeds surpassing even efficient local-descriptor methods like BPNN.

Our asynchronous GPU-CPU algorithm resolves memory constraints by decoupling NL construction on CPUs from force evaluation on GPUs. Crucially, standard implementations of GPU-based NL building in the JAX_MD package must satisfy JIT compilation constraints, incurring prohibitive memory overhead for general lattice parameters. By shifting NL operations to CPUs with optimized Fortran codes, we enable simulations of nearly one million atoms compared to the 50k-atom limit imposed by the JAX_MD implementation without any extra overhead. This is significant as the NEMP can be used to simulate much larger systems than permitted by the existing EEMPs.

The current computational bottleneck lies in the EC calculation, where an NN processes edge-invariant input features. Although the channel contractions in Eqs. (10) and (12) exhibit quadratic scaling, these operations remain subdominant as they only operate on the node dimensions. The EC evaluation, which scales linearly with the number of channels, can be further accelerated using TensorFloat-32 optimization. When combined with the parallel algorithms for MPNN frameworks, such as that proposed recently by Xia and Jiang[75], our approach can be further extended to larger-scale systems. The NEMP model's unique features make it particularly suited for advancing MD simulations and for the development of accurate, efficient universal potentials for chemical, biophysical, and materials studies at ab initio accuracy. Furthermore, the exceptional performance demonstrates that our strategy provides a simple and general approach to upgrading other EEMP models, improving efficiency in both computational time and memory usage.

**Methods**

**Loss function.** Networks are trained using a loss function based on a weighted sum of energy and forces:

$$\mathcal{L} = \lambda_E \sum_{i=1}^{N_b} \left( \frac{E_i^{\text{NN}} - E_i^{\text{DFT}}}{N_{atom}} \right)^2 + \lambda_F \sum_{i=1}^{N_b} \left\| \mathbf{F}_i^{\text{NN}} - \mathbf{F}_i^{\text{DFT}} \right\|^2. \tag{15}$$

Here $\lambda_E$ and $\lambda_F$ denote the energy and force weights, respectively, which are hyperparameters. $N_b$ is the batch size. We found that using a decaying force weight with learning rate decay—from approximately 10 to 0.1—while keeping the energy weight fixed at unity yields optimal test error. The only exception is 3BPA, for which we use a constant force weight of 0.2. $E^{\text{NN}}$ ($F^{\text{NN}}$) and $E^{\text{DFT}}$ ($F^{\text{DFT}}$) represent the NN–predicted energy (force) and DFT-calculated energy (force), respectively. Our package also supports training of the stress tensor; however, this feature was not employed in the present work and is therefore not discussed here.

**Training details**

**3BPA.** We employed an NEMP model with 3 MP layers, each having 64 channels. To improve efficiency in each MP layer, we used different maximal angular momenta: $L_{\max}$=3 (as in Eq. (5)) for evaluating spherical harmonics, and $l_{\max}$=2 (the maximal value of $l_f$ in Eq. (11)) for the tensor product in MP layer. The cutoff radius was set to 6.0 Å. The learning rate was initially warmed up from 0.01 to 0.2, then decayed by a factor of 0.5 whenever the validation loss did not decrease for 60 consecutive steps, until reaching a minimum of $1\times10^{-5}$. The batch size was set to 1. The NN structure of $F_{emb}$ comprised two ResNet blocks[56, 57], each with two layers of size 128×128 (denoted as 2×(128×128), used hereafter). Both $F_{coeff}$ and $F_{readout}$ layers were implemented as linear connections to enhance transferability.

**Liquid water.** We employed an NEMP model with 3 MP layers, each having 8 channels. Both $L_{max}$ and $l_{max}$ were set to 2. The cutoff radius was 4.5 Å. The learning rate was initially warmed up from 0.001 to 0.01, then decayed by a factor of 0.035 whenever the validation loss did not decrease for 150 consecutive steps, until reaching a minimum of $1\times10^{-5}$. The batch size was set to 1. The NN structure of the $F_{emb}$ was 2×(64×64), that of the $F_{coeff}$ was 1×(32×32), and that of the $F_{readout}$ was 2×(32×32). For the NEMP-small model, the MP layers were reduced to 2, each having 4 channels. The $F_{coeff}$ NN structure was reduced to 1×(16×16); all other settings remained the same as in the standard model.

**ANI-1x.** We employed an NEMP model with 3 MP layers, each having 32 channels. $L_{max}$ was set to 3 and $l_{max}$ were set to 2. The cutoff radius was 6.0 Å. The learning rate was initially warmed up from 0.001 to 0.01, then decayed by a factor of 0.5 whenever the validation loss did not decrease for 25 consecutive steps, until reaching a minimum of $1\times10^{-5}$. The batch size was set to 64. The NN structure of the $F_{emb}$ was 2×(128×128), that of the $F_{coeff}$ was 2×(128×128), and that of the $F_{readout}$ was 2×(256×256).

**HME21.** We employed an NEMP model with 3 MP layers, each having 16 channels. Both $L_{max}$ and $l_{max}$ were set to 2. The cutoff radius was 6.0 Å. The learning rate was initially warmed up from 0.001 to 0.01, then decayed by a factor of 0.035 whenever the validation loss did not decrease for 50 consecutive steps, until reaching a minimum of $1\times10^{-5}$. The batch size was set to 2. The NN structure of the $F_{emb}$ was $2\times(128\times128)$, that of the $F_{coeff}$ was $2\times(64\times64)$, and that of the $F_{readout}$ was $2\times(128\times128)$.

**EMLP.** We employed an NEMP model with 3 MP layers, each having 16 channels. Both $L_{max}$ and $l_{max}$ were set to 2. The cutoff radius was 5.0 Å. The learning rate was initially warmed up from 0.002 to 0.02, then decayed by a factor of 0.5 whenever the validation loss did not decrease for 20 consecutive steps, until reaching a minimum of $1\times10^{-6}$. The batch size was set to 32. The NN structure of the $F_{emb}$ was $2\times(128\times128)$, that of the $F_{coeff}$ was $2\times(64\times64)$, and that of the $F_{readout}$ was $2\times(128\times128)$.

**Data availability:**

All datasets used for model training and validation in this study are publicly available from previously published sources:

3BPA dataset (https://github.com/davkovacs/BOTNet-datasets);

liquid water dataset (https://github.com/BingqingCheng/ab-initio-thermodynamics-of-water);

ANI-1x dataset (https://springernature.figshare.com/articles/dataset/ANI-1x_Dataset_Release/10047041);

COMP6 dataset (https://github.com/isayev/COMP6/tree/master);

HME21 dataset (https://figshare.com/articles/dataset/High-temperature_multi-element_2021_HME21_dataset/19658538);

EMLP dataset (https://github.com/HuGroup-shanghaiTech/REICO).

**Code availability:**

The NEMP Package is available from https://github.com/zhangylch/NEMP.

**Acknowledgements**:

This work was supported by Department of Energy (Grant No. DE-SC0015997 to H.G.). The computation was performed at the Center for Advanced Research Computing (CARC) at UNM. We gratefully acknowledge Dr. Peijun Hu and Dr. Wenbo Xie for providing the EMLP dataset and the corresponding validation data for the potential.

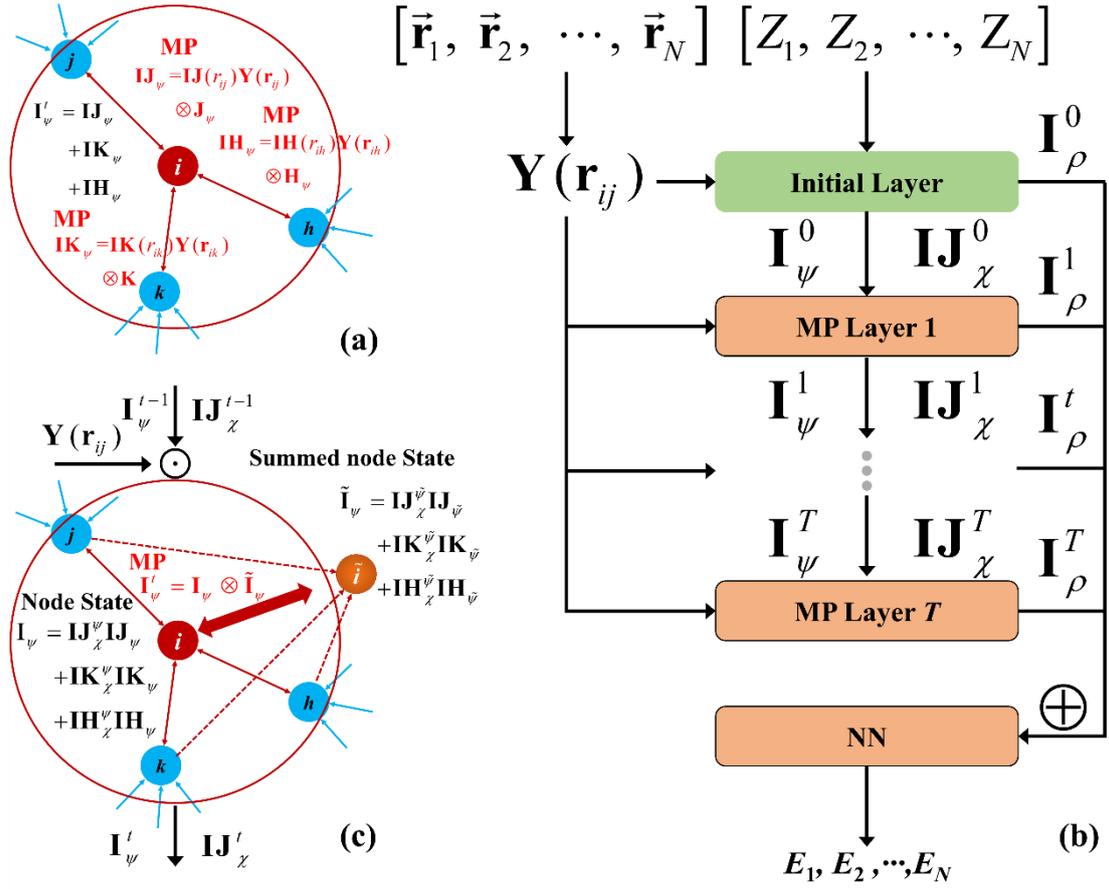

**Fig. 1 Message passing mechanisms and NEMP architecture.** Schematic illustrations of (a) the message passing mechanism for EEMP, (b) the NEMP architecture, and (c) the message passing mechanism for NEMP. Here, $\otimes$ denotes the tensor product, '$i$' represents the central node, and '$j, k, h$' denote the neighboring nodes.

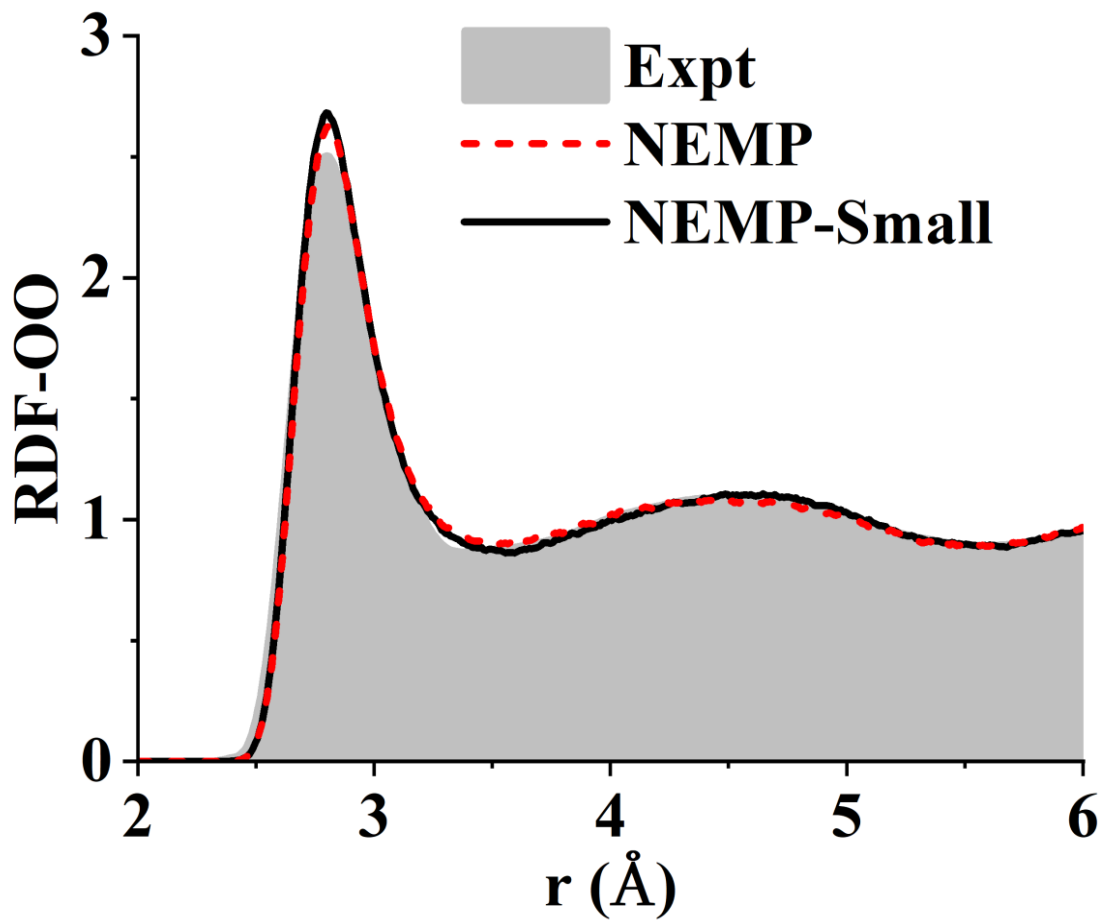

**Fig. 2 Radial distribution functions of liquid water.** Comparison of experimental[63] and theoretical O-O radial distribution functions of liquid water obtained by MD simulations with an NEMP and NEMP-Small potential at 300 K.

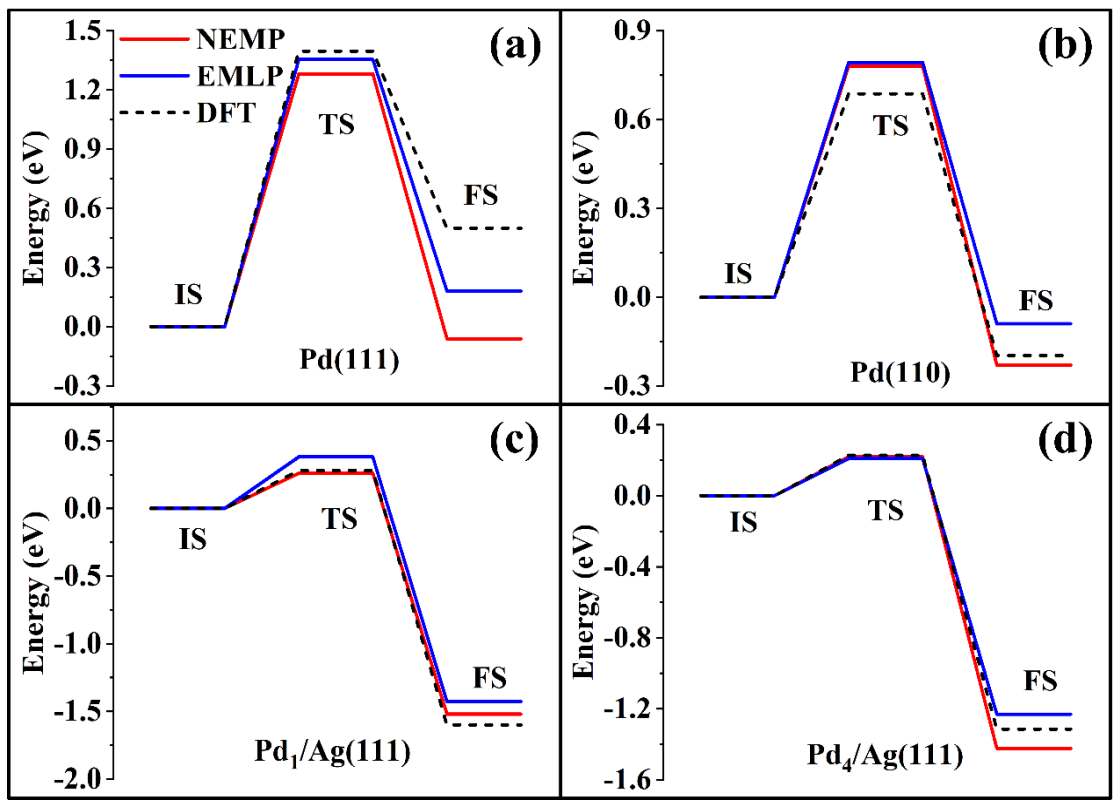

**Fig. 3 Reaction pathways for CO oxidation on Pd-based surfaces.** CO oxidation reaction path on (a) Pd(111), (b) Pd(110), (c) Pd(111), and (d) Pd$_4$/Ag(111), comparing DFT calculations with EMLP/NEMP predictions.

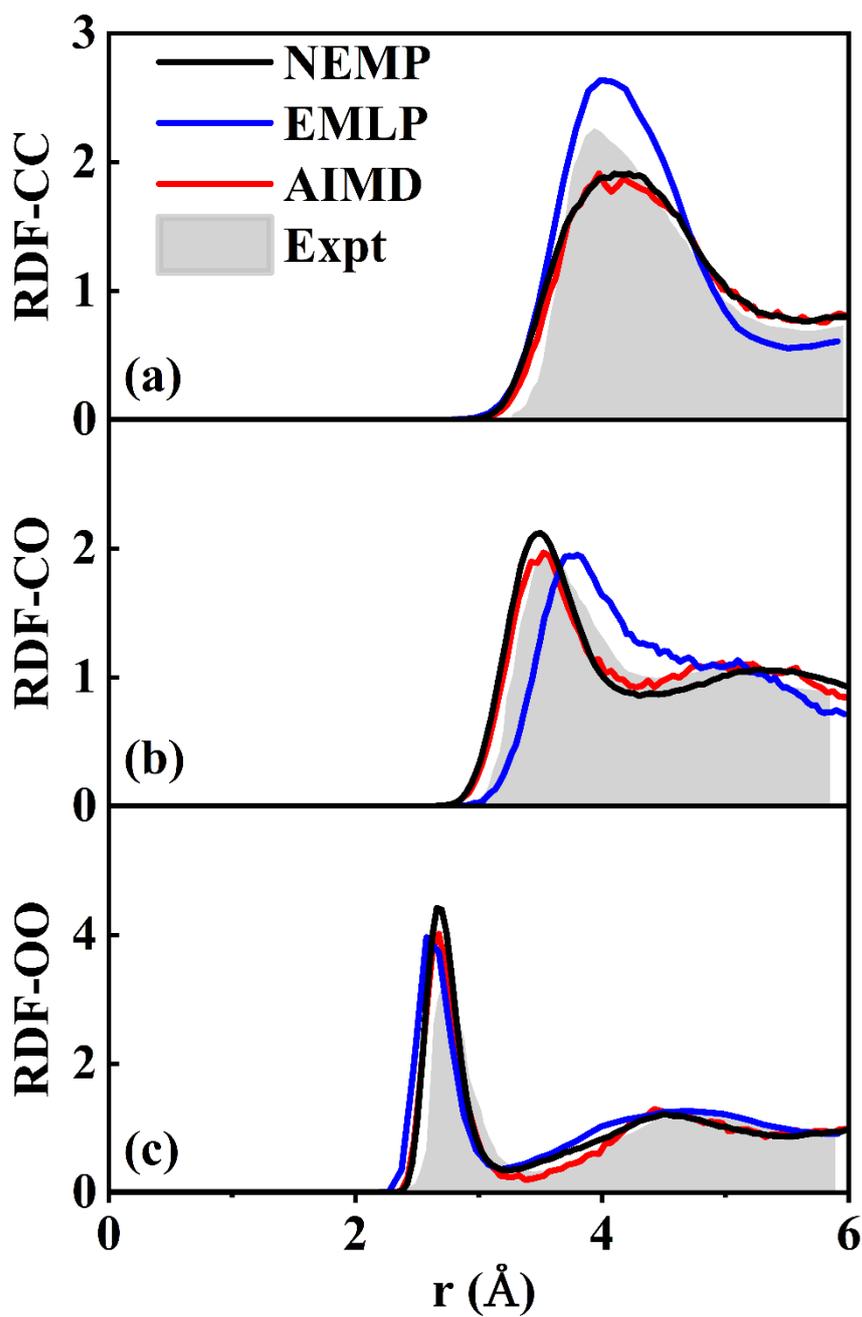

**Fig. 4 Radial distribution functions of liquid methanol.** Comparison of experimental[73] and theoretical RDFs of liquid methanol at 300 K for (a) C‑C, (b) C‑O, and (c) O‑O pairs, obtained using DFT, and EMLP, and NEMP potentials.

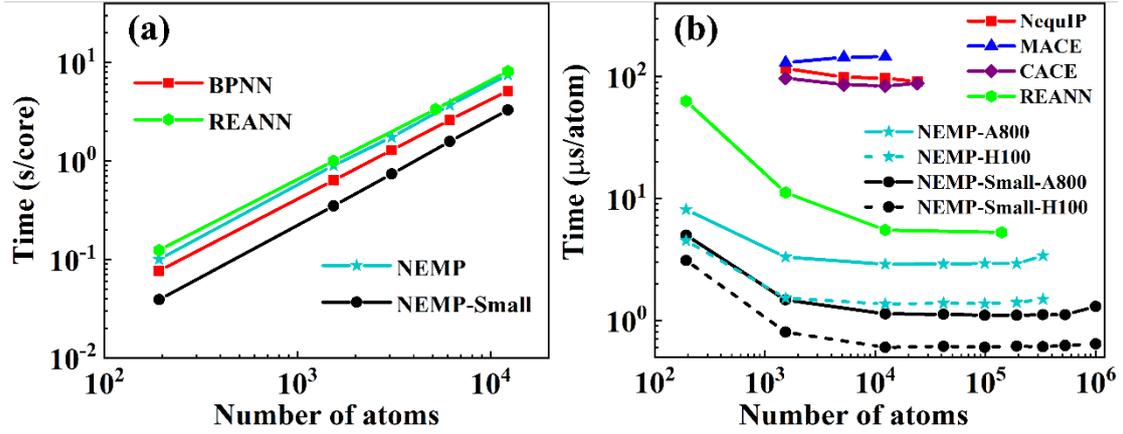

**Fig. 5 Computational cost per MD step in liquid water simulations.** Computational cost per MD step in liquid water simulations for: (a) NEMP and NEMP-Small, evaluated on a single core of Intel® Xeon Gold 6438Y CPUs and, compared with BPNN and REANN, whose results are taken from Refs. [57, 62], were computed on a single core of Intel® Xeon 6132 CPU; and (b) NEMP, NEMP-Small (A800/H100 GPUs) versus REANN, CACE, NequIP, and MACE (A100 GPUs). Note that all calculations of the NEMP model are conducted using float32 precision.

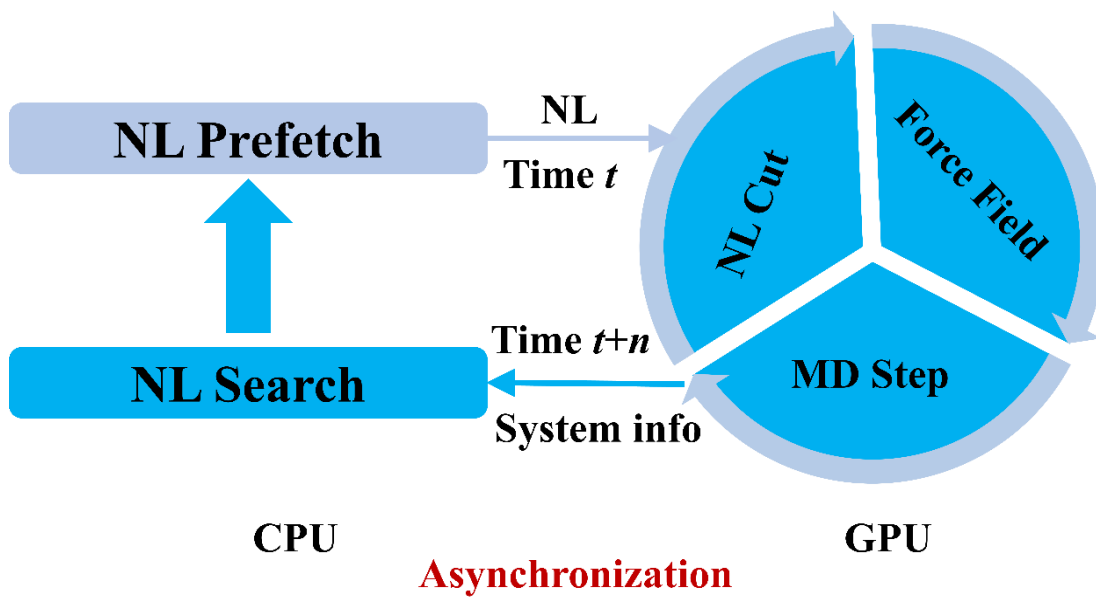

**Fig. 6 Asynchronous workflow of NL generation and MD propagation.** Schematic of the asynchronous computation workflow showing NL generation on CPU concurrent with MD propagation on GPU.

Table I: Prediction errors (RMSEs) for energy (meV) and atomic forces (meV/Å) of the 3BPA molecule across different MLIPs. The boldfaced number indicates the model with the lowest error. 'E' and 'F' in the first column denote energy and forces, respectively.

| Dataset | ACE | sGDML | ANI-2x | NequIP | CACE | MACE | NEMP |
|---|---|---|---|---|---|---|---|
| 300 K, E | 7.1 | 9.1 | 38.6 | 3.3 | 6.3 | **3.0** | 3.3 |
| 300 K, F | 27.1 | 46.2 | 84.4 | 11.3 | 21.4 | **8.8** | 10.5 |
| 600 K, E | 24.0 | 484.8 | 54.5 | 11.2 | 18.0 | 9.7 | **9.7** |
| 600 K, F | 64.3 | 439.2 | 102.8 | 27.3 | 45.2 | **21.8** | 24.4 |
| 1200 K, E | 85.3 | 774.5 | 88.8 | 40.8 | 58.0 | **29.8** | 33.1 |
| 1200 K, F | 187.0 | 711.1 | 139.6 | 86.4 | 113.8 | **62.0** | 67.9 |
| Time | \ | \ | \ | 103.5 | \ | 24.5 | **2.4** |

Table II: RMSEs for energy per water molecule (meV/$H_2O$) and atomic forces (meV/Å) across different MLIPs for liquid water. The boldfaced number indicates the model with the lowest error.

|        | BPNN  | EANN  | REANN | NequIP | CACE | MACE | NEMP |
|--------|-------|-------|-------|--------|------|------|------|
| Energy | 7.0   | 6.3   | 2.0   | 2.8    | **1.8** | 1.9  | 2.3  |
| Forces | 120.0 | 129.0 | 47.0  | 45.0   | 47.0 | **36.2** | 37.3 |

Table III: Performance comparison of MAEs for total energies (kcal/mol) and atomic forces (kcal/mol/Å) across different MLIPs for the COMP6 benchmark dataset. Note that the ANI-1x model utilized 10× more training data than other models. The boldfaced numbers indicate the model with the lowest error. 'E' and 'F' in the first column denote energy and forces, respectively.

| Dataset | ANI-1x | TensorNet | MACE | **NEMP** |
|---|---|---|---|---|
| ANI-MD E/F | 3.4/2.68 | **1.61**/0.82 | 3.25/0.62 | 2.86/**0.46** |
| DrugBank E/F | 2.65/2.86 | 0.98/0.75 | 0.73/0.47 | **0.65/0.34** |
| GDB 7–9 (E) | 1.04/2.43 | 0.32/0.53 | **0.21**/0.34 | 0.24/**0.24** |
| GDB 10–13 (E) | 2.3/2.67 | 0.83/1.52 | **0.53**/0.62 | **0.53/0.46** |
| S66x8 (E) | 2.06/1.6 | 0.62/0.33 | **0.39**/0.22 | 0.45/**0.17** |
| Tripeptides (E) | 2.92/2.49 | 0.92/0.62 | 0.79/0.44 | **0.67/0.34** |
| COMP6 total (E) | 1.93/2.09 | \ | **0.48**/0.52 | **0.48/0.39** |

Table IV: MAEs for Energy per atom (meV/atom) and Force (meV/Å) of several MLIPs on the HME21 dataset. The boldfaced number indicates the model with the lowest error.

|        | TeaNet | NequIP | MACE  | NEMP  |
|--------|--------|--------|-------|-------|
| Energy | 19.6   | 47.8   | 16.5  | **16.2** |
| Forces | 175    | 199    | **140.2** | 144.3 |